\documentstyle[12pt]{article}
  \advance\oddsidemargin  by -1in
  \advance\evensidemargin by -1in
  \advance\topmargin	  by -0.5in
\def\lsim{\mathrel{\rlap{\lower4pt\hbox{\hskip1pt$\sim$}}
    \raise1pt\hbox{$<$}}}	  
\def\gsim{\mathrel{\rlap{\lower4pt\hbox{\hskip1pt$\sim$}}
    \raise1pt\hbox{$>$}}}	  

\def\beq{\begin{equation}}
\def\endeq{\end{equation}}
\def\arr{\begin{eqnarray}}
\def\endarr{\end{eqnarray}}
\makeindex

\begin{document}

\begin{center}
{\Huge \bf
Novel color transparency effect: \\
scanning the wave function of vector mesons \vspace{1.5cm}\\ }
{\Large B.Z.Kopeliovich$^{1,2)}$, J.Nemchick$^{2,3)}$, N.N.Nikolaev$^{4)}$\\
and B.G.Zakharov$^{4)}$\bigskip\\ }
{\sl
$^{1)}$TRIUMF,4004 Wesbrook Mall, Vancouver, B.C. V6T 2A3, Canada\\
$^{2)}$Laboratory of Nuclear Problems\\
Joint Institute for Nuclear Research\\
Head Post Office, P.O.Box 79, 101000 Moscow, Russia\\
$^{3)}$Institute of Experimental Physics SAV,\\
Solovjevova 47, CS-04353 Kosice, Slovakia\\
$^{4)}$L.D.Landau Institute for Theoretical Physics, GSP-1, 117940,\\
ul.Kosygina 2, V-334, Moscow, Russia\\}

\bigskip
\bigskip
\bigskip

{\bf \Large A b s t r a c t}
\end{center}

We demonstrate how the virtual photoproduction
of vector mesons on nuclei scans the wave function of
vector mesons from the large non-perturbative
transverse size $\rho \sim R_{V}$ down to the small
perturbative size  $\rho \sim 1/\sqrt{Q^{2}}$.
Thee mechanism of scanning is based on color transparency and QCD
predicted spatial wave function of quark-antiquark fluctuations
of virtual photons.
A rich, energy- and $Q^{2}$-dependent, pattern of
the nuclear shadowing and antishadowing is predicted,
which can be tested at the European Electron Facility
and SLAC.	\vspace{1.5cm}\\
\pagebreak


{\bf Introduction.}
In this paper we discuss the novel feature of color transparency (CT) tests
of QCD: the scanning of the nonperturbative
hadronic wave functions. The
virtual photoproduction of vector mesons ($\Upsilon,\,
\Upsilon',\, J/\Psi,\,\Psi',\, \rho,\,\rho',...$)
is particularly suited for such CT tests. At high energy $\nu$,
the photoproduction can be viewed [1,2] as a production of
the virtual $\bar{q}q$ pair with the coherence length
\begin{equation}
l_{c}={{2\nu} \over Q^{2}+m_{V}^{2}} \, .
\label{eq.1}
\end{equation}
The size $\rho_{Q}$ of the produced $\bar{q}q$ pair (the
ejectile state) is controlled by the virtuality $Q^{2}$ of
photons [3].
The $\bar{q}q$ state is projected onto (recombines into) the
final-state vector meson $V$ with the formation (recombination)
length
\begin{equation}
l_{f}={\nu \over m_{V}\Delta m} \, ,
\label{eq:2}
\end{equation}
where $\Delta m $ is the typical level splitting in the
quarkonium. We demonstrate, how by changing $Q^{2}$ and
the initial size $\rho_{Q}$ one
can scan the wave function of vector mesons starting
from the non-perturbative size $\rho \sim R_{V}$
down to the perturbative region of $\rho \sim 1/Q$.

Besides the scanning radius $\rho_{Q}$, the formation length
$l_{f}$ is a second important parameter of CT physics.
Changing energy $\nu$, one can vary $l_{f}$
from $l_{f}\ll R_{A}$ (quasi-instantaneous
formation of the final-state hadron) to $l_{f} > R_{A}$
(the frozen-size limit), and thus study
the dynamical evolution of the small-sized, perturbative
$q\bar{q}$ pair to the full-sized nonperturbative hadron.
We predict very rich pattern of the nuclear shadowing
and antishadowing phenomena, which changes with the scanning
radius $\rho_{Q}$, i.e., with $Q^{2}$, and with the evolution
rate, i.e., with energy $\nu$.

It is worth while to emphasize that the virtual photoproduction
exemplifies in a particularly transparent way the principle
idea of CT tests of QCD:
(1) A small transverse-size component of the interacting
hadrons, $\rho \ll R_{h}$, is selected by the interaction
dynamics. (2) The interaction cross section $\sigma(\rho)$
of the small-sized ejectile is measured by a strength of
the final state interaction
(FSI) in the target nucleus [4,5]. Notice a combination of
the {\sl perturbative} and {\sl nonperturbative} aspects of QCD:
the {\sl perturbative} production of the small-sized
ejectile is followed by probing its size in the
{\sl nonperturbative}, diffractive small-angle scattering
in the nuclear matter. Nevetheless, QCD as a theory of strong
interactions predicts that the
strength of this diffractive scattering vanishes as
$\rho \rightarrow 0$: $\sigma(\rho)\propto \rho^{2}$ [6,7].
\smallskip\\


{\bf The scanning mechanism.}
The quantum-mechanical description of the scanning goes as
follows:
At $l_{f},\,l_{c} > R_{A}$ the amplitude of the forward
photoproduction on the free nucleon $M_{N}$ and the nuclear
transmission coefficient or the nuclear transparency
$Tr_{A}=d\sigma_{A} /Ad\sigma_{N}$ in the quasielastic
photoproduction $\gamma^{*}A\rightarrow VA$ read [8,9]:
\begin{equation}
M_{N}=\langle V|\sigma(\rho)|\gamma^{*}\rangle
\label{eq.3}
\end{equation}
\begin{equation}
Tr_{A}={1\over A}
\int d^{2}\vec{b} T(b)
{\langle V |\sigma(\rho)
\exp\left[-{1\over 2} \sigma(\rho)T(b)\right] |\gamma^{*}
\rangle^{2} \over
\langle V|\sigma(\rho)|\gamma^{*}\rangle^{2} }
\label{eq:4}
\end{equation}
where $T(b)=\int dz n_{A}(b,z)$ is the optical thickness
of a nucleus, the nuclear density $n_{A}(b,z)$ is
normalized to the nuclear mass number A:
$\int d^{3}\vec{r} n_{A}(\vec{r}) = A$.

The wave function $|\gamma^{*}\rangle$
of the $q\bar{q}$ fluctuations of the virtual photons
was calculated in [3] in the mixed $(\vec{\rho},z)$
representation, where $z$ is a fraction of photon's
light-cone momentum, carried by the quark, $0 <z<1$.
The most important feature of
$|\gamma^{*}\rangle$ is an exponential decrease
at large distances [3]:
\begin{equation}
|\gamma^{*}\rangle
\propto \exp(-\varepsilon \rho)
\label{eq.5}
\end{equation}
where
\begin{equation}
\varepsilon^{2} = m_{q}^{2}+z(1-z)Q^{2}
\label{eq.6}
\end{equation}
Calculation of the matrix elements in (3,4) involves the
$d^{2}\vec{\rho} dz$ integration. In the nonrelativistic
quarkonium $z \approx 1/2$, so that the relevant $q\bar{q}$
fluctuations have a size
\begin{equation}
\rho \sim \rho_{Q} = {1 \over \sqrt{m_{q}^{2}+{1\over 4}Q^{2}}}
\approx {2 \over \sqrt{m_{V}^{2}+Q^{2}}}
\label{eq:7}
\end{equation}

What enters (3,4) is a product $\sigma(\rho)|\gamma^{*}\rangle$.
Because of CT property,
$\sigma(\rho) \propto \rho^{2}$ at small $\rho$ [6,7], this product
will be sharply peaked at
$\rho \approx \rho_{2}=2\rho_{Q}$ with the
width $\Delta \rho \sim 2\rho_{Q}$, which leads naturally to
the idea of scanning: The transition
matrix elements  (3,4) probe the wave function of vector
mesons at $\rho \sim 2\rho_{Q}$, and varying $\rho_{Q}$
by changing $Q^{2}$, one can
scan the wave function $|V\rangle$ from large to small
distances. In Fig.1 we demonstrate qualitatively how the
scanning works. We also show the $z$-integrated
wave functions of the ground state $|V\rangle$ and
of the radial excitation $|V'\rangle$.

The nuclear matrix element in (4) can be expanded in the moments
$\langle V| \sigma(\rho)^{n}|\gamma^{*}\rangle $
which are the true {\sl QCD observables} of the
virtual photoproduction process. These moments probe the
wave function $|V\rangle$ at different values of $\rho
\sim \rho_{2n}$. To the leading
order in the final state interaction,
\begin{equation}
Tr_{A}=1-\Sigma_{V} {1\over A}\int d^{2}\vec{b}T(b)^{2}
\label{eq.8}
\end{equation}
where
\begin{equation}
\Sigma_{V}={\langle V|\sigma(\rho)^{2}|\gamma^{*}\rangle
\over \langle V|\sigma(\rho)|\gamma^{*}\rangle }
\label{eq.9}
\end{equation}
 This expansion works well when $1-Tr_{A} \ll 1$, and is
convenient to explain how the scanning proceeds.
Notice, that $\sigma(\rho)^{2}|\gamma^{*}\rangle$
peaks at $\rho \sim \rho_{4}=4\rho_{Q}$.\smallskip\\


{\bf Scanning the vector mesons: the node effect}.~~
Start with the real photoproduction ($Q^{2}=0$) of the
ground-state vector mesons
($\Upsilon,\, J/\Psi,\,\rho^{0},...$) and take the case
of the charmonium. In this case $\rho_{Q} =
1/m_{c} \ll R_{J/\Psi}$, \,\, $\rho_{2}$ is
still smaller than $R_{J/\Psi}$, but $\rho_{4}\sim
R_{J/\Psi}$. For this numerical reason, one finds
$\Sigma_{J/\Psi} \approx \sigma_{tot}(J/\Psi\,N)$, i.e.,
the predicted nuclear shadowing will be
marginally similar [8,9] to the Vector Dominance Model
(VDM) prediction [10]. This holds to much extent for the
$\Upsilon$ and the light vector mesons as well.
At larger $Q^{2}$, one has $\rho_{2},\,\rho_{4}
\ll R_{V}$ and
$\Sigma_{V} \sim \sigma(\rho_{Q}) \propto \rho_{Q}^{2}$
with the calculable logarithmic corrections [3,11], so that
the above marginal similarity with the VDM disappears.
The nuclear transparency will tend to unity from below:
\begin{equation}
1-Tr_{A} \propto {A \over R_{A}^{2}}\rho_{Q}^{2}
\label{eq.10}
\end{equation}

The case of the radial excitation  $V'$ is more interesting.
Radial excitations have larger radius and larger
free nucleon cross section: e.g., $\sigma_{tot} (\Psi' N)
\approx 2.5 \sigma_{tot}(J/\Psi \,N)$ [8], which classically
would
suggest much stronger final state interaction for $\Psi'$
than for $J/\Psi$. Presence of the node in the $V'$ wave
function leads to a rather complex pattern of
the shadowing and antishadowing. In the photoproduction
limit, because of $\rho_{2} \sim R_{\Psi'}$, there are
rather strong cancellations between the contributions
to the amplitude (3) from $\rho$ below and above the
node (the node effect, see Fig.1).
For this reason, in photoproduction on the free
nucleons, one predicts the ratio of the forward production
differential cross sections
$r(Q^{2}=0) =d\sigma(\gamma N\rightarrow V'N)/
d\sigma(\gamma N \rightarrow VN)|_{t=0} < 1$.
For the charmonium the $\Psi'/(J/\Psi)$ ratio
$r(0) = 0.17$.
In the $\Upsilon'$ production, the initial size
$\rho_{Q}$ scales as $1/m_{q}$, whereas the radius of
the $\bar{b}b$ bound states decreases with $m_{q}$ less
rapidly. For this reason for the $\Upsilon'$ the node
effect is much weaker and we find the
$\Upsilon'/\Upsilon$ ratio $r(0)=0.84$.
The larger is $Q^{2}$, the smaller size $\rho_{2}$
is scanned, a contribution
from the region above the node becomes negligible,
and $r(Q^{2})$ will increase with $Q^{2}$ up to
$r(Q^{2} \gg m_{V}^{2}) \sim 1$ (the exact limiting
ratio depends on the wave function's at the origin).

Since $\rho_{4}$ is closer to the node position, the
node effect is still stronger in the matrix element
$\langle V'|\sigma(\rho)^{2} |\gamma\rangle$.
For the $\Psi'$ one finds
$\langle \Psi'|\sigma(\rho)^{2} |\gamma\rangle <0$, so that
$\Sigma_{\Psi'} <0$ and  $Tr_{A}(\Psi') >1$
despite the larger free-nucleon cross section [8,9,11].
In the $\Upsilon'$ production
the scanning radius compared to the bottonium radius is
relatively smaller than in the charmonium case,
the node effect is weaker and FSI produces the shadowing
of the $\Upsilon'$. Still, in spite of
 $\sigma_{tot}(\Upsilon'N)\gg  \sigma_{tot}(\Upsilon N)$,
 shadowing of the $\Upsilon'$ is weaker than shadowing of
the $\Upsilon$.

The $Q^{2}$-dependent scanning changes the
node effect significantly. The smaller the scanning radius
$\rho_{Q}$, the weaker are the cancellations. In the $\Psi'$
case $\langle V'|\sigma(\rho)^{2}|\gamma^{*}\rangle$
becomes positive defined, so that the antishadowing changes
to the shadowing, which first rises with $Q^{2}$, then
saturates and is followed by the onset of asymptotic decrease (10).
In the $\Upsilon'$ production the node effect is weaker,
still it affects the $Q^{2}$-dependence of the shadowing
making it different from that of the $\Upsilon$.

Predictions for the heavy quarkonium photoproduction are
shown in Figs.2,3. Because of the small size of heavy
quarkonium, the results are numerically reliable, as they are
dominated by the perturbative QCD domain.  \smallskip\\

{\bf The two scenarios of scanning the light vector mesons.}
The case of the light vector mesons is particularly interesting,
as similar (anti)shadowing effects occur in the energy and
$Q^{2}$ range
accessible at SLAC and at the planned European Electron
Facility (EEF).
Here at moderate $Q^{2}\lsim m_{\rho}^{2}$ the scanning radius
is large, $\rho_{Q} \sim R_{V}$, so that numerically the
predicted shadowing of the $\rho^{0}$ , Fig.4, is not very accurate,
although the accuracy increases gradually with $Q^{2}$ as
the scanning radius decreases into the perturbative domain
$\rho_{Q} \ll R_{V}$. Nevetheless, since the wave function of the
$\rho^{0}$ does not have a node, we beleive to describe correctly
a smooth transition from the real to virtual photoproduction.

In the photoproduction limit we find a strong node
effect and strong suppression of the
photoproduction of the radial excitation $\rho'$  on nucleons,
by more than one order of magnitude
compared to the $\rho^{0}$ production, which broadly agrees
with the scanty experimental data [10,12]. As here
$\rho_{Q} \sim R_{V}$, we can not give a reliable numerical
estimate of how small the $\rho'/\rho^{0}$ cross section
ratio is. Nevertheless, we can describe the two
possible scenarios of scanning the $\rho'$
wave function, experimental tests of which can shed light on
the spectroscopy and identification of the radial
excitations of the light vector mesons:\\
{\bf
(1)
The
undercompensated
free-nucleon amplitude}:
$\langle \rho'|\sigma(\rho)|\gamma\rangle  >0$.\\
In this scenario the $\rho'$ case will be similar
to the $\Psi'$ case, apart from the possibility of
anomalously strong nuclear enhancement $Tr_{A} \gg 1$,
which might show strong atomic number dependence.
Indeed, because of the	larger relevant
values of $\rho$ the cross section $\sigma(\rho)$ is
larger, and the
attenuation factor $\exp[-{1 \over 2}\sigma(\rho)T(b)]$
in the nuclear matrix element, eq.(4), will suppress
the large $\rho$ region, effectively decreasing the
scanning radius $\rho_{Q}$ and diminishing the
node effect. Detailed description of
the atomic number dependence will be presented elsewhere.

The $Q^{2}$-scanning too will follow the $\Psi'$-scenario:
change from the antishadowing to the shadowing with increasing
$Q^{2}$, followed by the saturation and then decrease of the
shadowing according to eq.(10). The range of $Q^{2}$ at
which the major effects should occur corresponds to a
change of the scanning radius
$\rho_{Q}$ by the factor $\sim 2$, i.e., to
$Q^{2} \sim 4m_{q} \sim m_{\rho}^{2}$. \\
{\bf
(2)
The
overcompensated free-nucleon
amplitude}: $\langle \rho'|\sigma(\rho)|\gamma\rangle < 0$.\\
This scenario is preferred in the crude oscillator model
used in [3] and fits the pattern of the node effect becoming
stronger for the lighter flavours.
In this case $\rho_{2} \gsim R_{V}$ and
the higher moments too will be negative
valued:
$\langle \rho'|\sigma(\rho)^{n}|\gamma\rangle < 0$, so that
in the photoproduction limit one starts with the
conventional shadowing: $Tr_{A} < 1$.

In the $Q^{2}$-scanning process the striking
effect is bringing the free-nucleon amplitude
$\langle V'| \sigma(\rho)|\gamma^{*}\rangle$ to
the exact compensation at certain moderate $Q^{2}$,
when the decreasing $\rho_{2}$ intercepts $R_{V}$
(strictly speaking, because of the relativistic
corrections and different quark
helicity states, the compensation is unlikely to be
exact). As a result, one finds a spike in $Tr_{A}$,
Fig.3.
With the further increase of $Q^{2}$ one enters the
undercompensation regime, and the futher pattern
of the $Q^{2}$-scanning will be essentially
the same as in the undercompensation or the $\Psi'$
scenario.

The above described $Q^{2}$-dependent scanning of the
wave function of light vector
mesons ofers a unique possibility of identifying the
radial excitation of the light vector mesons (for
the detailed discussion of the spectroscopy of light
vector mesons see [12]). The corresponding
experiments could easily be performed at SLAC and
EEF. \smallskip\\


{\bf Quantum evolution, coherence and energy dependence of FSI.}

If the coherence length $l_{c} > R_{A}$, then
amplitudes of production on different nucleons add up coherently.
At moderate energy, $l_{c} < R_{A}$, the
production rates on different nucleons at the same impact
paramater $\vec{b}$ add up incoherently. In the opposite
limit of $l_{c} > R_{A}$ amplitudes of production on
different nucleons add up coherently, and nuclear affects
are generally weaker [8,9,11].

For the heavy quarks we have
a strong inequality $l_{f} \gg l_{c}$ [13]. The same inequality
holds for the light vector mesons at $Q^{2} \gg m_{V}^{2}$.
Consequently, at moderate energy, when $l_{c} < R_{A}$,
one has still a broad energy range in which  $l_{f} > R_{A}$
and the transverse
size of the $q\bar{q}$ pair is still frozen. In this
regime the nuclear transparency
is given by a simple formula
\begin{equation}
Tr_{A}={1 \over A}\int d^{2}\vec{b}dz n_{A}(b,z)
{ \langle V|\sigma(\rho)\exp[-{1 \over 2}\sigma(\rho)t(b,z)]
|\gamma^{*}\rangle ^{2}
\over
\langle V|\sigma(\rho)|\gamma^{*}\rangle^{2} }
\label{eq.11}
\end{equation}
where $t(b,z)=\int_{z}^{\infty}dz'n_{A}(b,z')$.
Notice, that compared to the high-energy limit (4), here the
attenuation effect in the nuclear matrix element is weaker.
The energy dependence
of the nuclear transparency is given by an approximate interpolation
formula
\begin{equation}
Tr_{A}(\nu)\approx Tr(l_{c} << R_{A})+F_{ch}(\kappa)^{2}
[Tr_{A}(l_{c} >R_{A})-Tr_{A}(l_{c} <<R_{A})]
\label{eq.12}
\end{equation}
where $F_{ch}(\kappa)$ is the charge form factor of the target
nucleus and $\kappa =1/l_{c}=(Q^{2}+m_{V}^{2})/2\nu$.
(For the general idea of derivation of (14) and
successful description of the NMC data [14] of the photoproduction
of the $J/\Psi$ see Ref.9).

If $l_{f} < R_{A}$, the spatial expansion of the $q\bar{q}$ pair
becomes important. In can be described
in either the quark basis used above, or in the hadronic
basis, which is nice demonstration of the quark-hadron deuality.

Consider the leading term of the final state
interaction in eq.(10) in the hadronic basis, i.e.,
in terms of Gribov's inelastic shadowing [15].
Inserting a complete set of the intermediate states, one can write down
\begin{equation}
\langle V|\sigma(\rho)^{2}|\gamma^{*}\rangle =
\sum _{i} \langle V|\sigma(\rho)|V_{i}\rangle
\langle V_{i} |\sigma(\rho)|\gamma^{*}\rangle
\label{eq.13}
\end{equation}
In the hadronic basis the
antishadowing of the $\Psi'$ comes from the desctructive
interference of the direct, VDM-like rescattering
\begin{equation}
\gamma^{*} \rightarrow \Psi' \rightarrow \Psi'
\label{eq.14}
\end{equation}
and the off-diagonal rescattering
\begin{equation}
\gamma^{*} \rightarrow J/\Psi \rightarrow Psi'
\label{eq.15}
\end{equation}
(there is a small contribution from
other intermediate states too).
The reason for the strong cancellation is that
the $\gamma \rightarrow \Psi'$ transition is weak
compared to the $\gamma \rightarrow J/\Psi$ transition,
whereas the $J/\Psi \rightarrow \Psi'$ transition
has an amplitude of opposite sign, and smaller, that
the $\Psi' \rightarrow \Psi'$ elastic scattering amplitude.
To the contrary, in the $J/\Psi$ photoproduction, both
amplitudes in the off-diagonal transitions like
$\gamma \rightarrow \Psi' \rightarrow J/\Psi$ are small,
and this explains why one finds a marginal similarity
to the VDM in the photoproduction.
At larger $Q^{2}$ the node effect is no longer effective
in suppressing the $\gamma^{*} \rightarrow \Psi'$ transition,
the off-diagonal amplitudes become significant for the
$J/\Psi$ photoproduction too and one finds strong
departure from the VDM for the $J/\Psi$ too.

By the nature of CT experiments, at finite energy $\sigma(\rho)^{2}$
in the {\sl l.h.s.} of eq.(13) is the nonlocal operator:
the $q\bar{q}$ state produced on one nucleon is probed by
a second nucleon a distance $\Delta z \sim R_{A}$ apart.
As a result, the diagonal and off-diagonal
rescattering amplitudes acquire the relative phase [15,16]
$\Delta \varphi_{21} = \Delta z(m_{2}^{2}-m_{1}^{2})/2\nu
\sim \Delta z /l_{f}$. Upon the
integration over $\Delta z$ only the intermediate
states $|i\rangle$ for which $\Delta \varphi_{i1} \lsim 1$,
or
\begin{equation}
|m_{i}^{2}-m_{V}^{2}| < 2\nu /R_{A}
\label{eq.16}
\end{equation}
will contribute
to the {\sl r.h.s.} of eq.(13), so that the strength
of the final state interaction will change rapidly
from the near-threshold energy of $l_{f} \ll R_{A}$ to
a higher energy of $l_{f} > R_{A}$. We describe the
energy-dependence in this region using the quark-basis
path-integral
technique suggested in [3].

At $l_{f}<<R_{A}$ the amplitudes of transitions (16) and
(17) do not interfere. However, since they are of
comparable magnitude, for the $\Psi'$ the VDM prediction
for the nuclear shadowing breaks down even near the production
threshold: $Tr_{A}$ is larger than the Glauber model
prediction. For the $J/\Psi$ a similar incoherent
contribution is small at small $Q^{2}$, rises with $Q^{2}$
as described above, and the near-threshold value of $Tr_{A}$
rises too. In the $\Psi'$ case the dominant effect of the
$Q^{2}$-dependent scanning is that the
$\gamma^{*}\rightarrow \Psi'$ transition amplitude increases
with $Q^{2}$ relative to the $\gamma^{*} \rightarrow
J/\Psi$ amplitude, which enhances the diagonal (shadowing)
rescattering contribution compared to the off-diagonal
(antishadowing) rescattering contribution. As a result of
this competition the near-threshold value of $Tr_{A}(\Psi')$
first decreases significantly with $Q^{2}$, followed by
the $J/\Psi$-like behaviour at larger $Q^{2}$ (Fig.2).
For the both $J/\Psi$ and $\Psi'$ the near threshold
behaviour of $Tr_{A}$ is partially due to the kinematical
rise of the threshold energy with $Q^{2}$.
In the bottonium the intial size $\rho_{Q}$ compared to
the bottonium radius is relatively smaller, than in the
charmonium, and $Tr_{A}(\Upsilon')$ exhibits a monotonous
$Q^{2}$ and $\nu$ dependence (Fig.2).

With the rising energy the destructive interference
of transitions (14) and (15) leads to a rapid rise of
the nuclear transparency $Tr_{A}$ and the onset of
the antishadowing of the $\Psi'$ in the photoproduction limit.
At the larger $Q^{2}$ the smaller $\rho$ are scanned,
and similar rise of $Tr_{A}$ with energy ends up in the
shadowing region.

In the photoproduction of the light vector mesons
$l_{f} \sim l_{c}$ and the more elaborated technique
like the effective diffraction operator technique
developed in [17] is called upon. This will be a
subject of the further investigation. The effect on
the $Q^{2}$-dependence is not significant, though,
and the cited results for the photoproduction of
the $\rho^{0}$-mesons, Fig.4, were obtained still
assuming $l_{c}>l_{f}$. At large $Q^{2}$, when for all
vector mesons $l_{f} \gg l_{c}$, the nuclear
transparency for $\rho^{0},J/\Psi,\Upsilon$  exhibits
a similar $Q^{2}$- and $\nu$-dependence. \smallskip\\


{\bf Coherence effects and scaling law for FSI.}
According to eq.10, at large $Q^{2}$ the FSI effect scales
as $A^{1/3}/Q^{2}$. However, this scaling
law, suggested in [18], is valid only at the asymptotic
energy [16].

Indeed, at moderate energy, the coherence constraint
(16) limits the number of the interfering states $N_{eff}$.
To a crude approximation, in this case the effective
attenuation will be controlled by not a small size
$\rho_{Q}$ of the initial $q\bar{q}$ pair, but rather by
the least possible size $\rho_{min}$ of the wave packet,
which can be constructed on a truncated basis of $N_{eff}$
conspiring states [17].( Above we already have encounterd
a strong effect of the coherence on the energy-dependence
of $Tr_{A}$ at fixed $Q^{2}$.)
Only in the very high energy limit,
when $N_{eff}$ is not bounded from above,
$\rho_{min} \sim \rho_{Q}$. The constraint (16) is
less stringent for the heavy quarkonia, as the level
splitting $\Delta m \ll 2m_{q}$, and is more noticeable
for the light vector mesons, as here the $m^{2}$-splitting
between higher excitations, which enters the coherence
condition (16), rises rapidly with the mass.
In Fig.5 we present our
results for the $Q^{2}$ dependence of $1-Tr_{A}(\rho^{0})$
at fixed energy $\nu$: it is much weaker than $\propto
1/(Q^{2}+m_{V}^{2})$, which is an appropriate
variable in biew of eq.(7). At asymptotic energy the
scaling law (11) works for all vector mesons, Fig.5. A somewhat
late onsent of the scaling law (11) for the $\rho^{0}$ can
be understood in terms of the overcompensation scenario in
the  $\rho^{0},\rho'$ photoproduction, as one needs a
relatively larger $Q^{2}$ to make the node effect
negligible. \medskip\\

{\bf Conclusions:}
We have shown that QCD observables of CT experiments correspond
to scanning the non-perturbative hadronic wave functions with
the $Q^{2}$-dependent scanning radius $\rho_{Q}$. The strength
of the final state interaction in the exclusive virtual
photoproduction of vector mesons is shown to depend strongly
on the nodal structure of wave functions.
At large energy and
$Q^{2}$  we predict the universal pattern of shadowing
for all the photoproduced vector mesons. The node effect in
the virtual photoproduction can be used to identify the
radial excitation states.
The predicted $Q^2$ and energy dependence of virtual photoproduction
on nuclei can easily be tested at SLAC and planned European Electron
Facility. The dedicated experiments on virtual photoproduction of vector
mesons desrve special attention, since theoretical predictions of CT signal
are much more reliable, than in $(e,e'p)$ or $(p,p'p)$ reactions (for the
recent review see [19]).
\medskip\\

{\bf Acknowledgements}

B.Z.K. thanks Theory Division of TRIUMF for the hospitality.
\pagebreak

\pagebreak
{\bf Figure captions:}
\begin{itemize}
\item[Fig.1.] -
The qualitative pattern of the the $Q^{2}$-dependent scanning
of the wave functions of the ground state $V$ and the radial excitation
$V'$ of the vector meson. The scanning distributions
$\sigma(\rho)\Psi_{\gamma^{*}}(\rho)$ shown by the solid and dashed
curve have the scanning radii $\rho_{Q}$ differing by a factor 3.
All wave functions are in arbitrary units.

\item[Fig.2.] -
The predicted $Q^{2}$ and $\nu$-dependence of the nuclear
transparency in the virtual phtotoproduction of the heavy quarkonia.
The qualitative pattern is the same from the light to heavy nuclei.

\item[Fig.3.] -
The predicted $Q^{2}$-dependence of the nuclear transparency in the
virtual photoproduction of the $\Psi'$ (the undercompensation
scenario) and $\rho'$ (the overcompensation scenario) mesons.

\item[Fig.4.] -
The predicted $Q^{2}$ and $\nu-$dependence of the nuclear transparency
in the virtual photoproduction of the $\rho^{0}$-mesons.

\item[Fig.5.] -
Test of the scaling law (10) in the virtual photoproduction of the
$J/\Psi$ and $\rho^{0}$ mesons: \\
The right box - the $\rho^{0}$ production at fixed energy. The dashed
straight line corresponds to the $1/(Q^{2}+m_{\rho}^{2})$ dependence.\\
The left box - at the asymptotic energy,
the dotted straight line corresponds to the $1/(Q^{2}+m_{J/\Psi}^{2})$
dependence. \\

\end{itemize}
\end{document}